\documentclass[runningheads]{llncs}
\usepackage[T1]{fontenc}
\usepackage{graphicx}
\usepackage{amsmath}
\usepackage{subcaption}
\usepackage{graphicx}
\usepackage[nolist]{acronym}
\usepackage{cleveref}

\usepackage{listings}
\usepackage{xcolor}
\usepackage{float}

\usepackage{tabularx} 
\usepackage{booktabs} 

\begin{acronym}
\acro{llm}[LLM]{large language model}
\acro{cg}[CG]{conjugate gradient}
\end{acronym}
\begin{document}
\title{From Prompts to Performance: Evaluating LLMs for Task-based Parallel Code Generation}
\titlerunning{From Prompts to Performance}
%
\author{Linus Bantel\orcidID{0009-0009-0247-4748} \and
Moritz Strack \and
Alexander Strack\orcidID{0000-0002-9939-9044}
\and Dirk Pflüger \orcidID{0000-0002-4360-0212}}
\authorrunning{Bantel et al.}

\institute{Institute for Parallel and Distributed Systems, University of Stuttgart, Stuttgart, Germany\\
\email{st166660@stud.uni-stuttgart.de}\newline
\email{\{linus.bantel, alexander.strack, dirk.pflueger\}@ipvs.uni-stuttgart.de}}
\maketitle              
\begin{abstract}
\Acp{llm} show strong abilities in code generation, but their skill in creating efficient parallel programs is less studied.
This paper explores how \acp{llm} generate task-based parallel code from three kinds of input prompts: natural language problem descriptions, sequential reference implementations, and parallel pseudo code.
We focus on three programming frameworks: OpenMP Tasking, C\texttt{++} standard parallelism, and the asynchronous many-task runtime HPX.
Each framework offers different levels of abstraction and control for task execution.
We evaluate \ac{llm}-generated solutions for correctness and scalability.
Our results reveal both strengths and weaknesses of \acp{llm} with regard to problem complexity and framework.
Finally, we discuss what these findings mean for future \ac{llm}-assisted development in high-performance and scientific computing.
\keywords{Large Language Models  \and Code Generation \and Asynchronous Many-Task Systems \and OpenMP \and C\texttt{++} Standard Parallelism \and HPX}
\end{abstract}
\section{Introduction}
The rapid evolution of \acp{llm} has transformed software engineering by enabling automated code generation.
Yet their ability to produce correct and performant parallel programs remains largely unexplored.
Parallel code generation requires reasoning about data dependencies, workload decomposition, and synchronization.
Recent empirical studies show that current models perform substantially worse on parallel than on serial workloads, often producing non-scalable or inefficient implementations~\cite{Nichols_2024}.

Asynchronous task-based parallel programming provides a useful framework to examine these limitations, as it imposes more difficult parallelization problems compared to conventional fork-join approaches.
We focus on OpenMP Tasking, C\texttt{++} standard parallelism, and the asynchronous many-task runtime HPX, which span a wide range of abstraction levels and execution models.
We also investigate the impact of prompting input modality on model performance, as prior work shows that zero-shot parallelization remains largely unreliable and prone to concurrency errors~\cite{YADAV2025112543}. Specifically, we base our different input prompts on natural language, pseudo-code, and sequential implementations.

Furthermore, we evaluate how different classes of \acp{llm} generate asynchronous task-based code and analyze how prompt complexity and abstraction influence correctness and performance.
The remainder of this work is structured as follows.
\Cref{sec:related_work} surveys recent work on code generation capabilities in the HPC context.
\Cref{sec:methods} then describes the evaluation methodology.
The results are presented in \Cref{sec:results} and followed by the conclusion in \Cref{sec:conclusion}.

\section{Related Work}\label{sec:related_work}

The application of \acp{llm} to automatic code generation has rapidly evolved from general-purpose programming assistance to domain-specific high-performance computing (HPC) code.
Recent advances in foundation models such as ChatGPT~\cite{chatgpt4}, DeepSeek~\cite{deepseekv3}, and open-source alternatives including LLaMA-2~\cite{llama2} have demonstrated strong capabilities in generating syntactically correct and semantically meaningful code across a wide range of programming languages~\cite{code_gen_1,code_gen_2,code_gen_3,code_gen_4}.
These capabilities have sparked growing interest in using \acp{llm} to improve developer productivity in HPC applications, where parallel programming remains complex and error-prone.

Recent machine learning–based approaches have attempted to bridge the gap between parallelism detection and code synthesis.
AutoParLLM~\cite{autoparllm} represents an early attempt to combine graph neural network-based parallelism detection with \ac{llm}-guided code generation, highlighting the potential of hybrid approaches but also underscoring the challenges of correctness and generalization.
However, the most significant recent progress in parallel code generation has come from \ac{llm}-based methods.
Early evaluations of OpenAI Codex showed that \acp{llm} can generate HPC kernels using OpenMP, OpenACC, CUDA, and MPI constructs directly from natural language prompts or partial code, with varying success depending on training data availability~\cite{parallel_kernels}.
Subsequent work expanded these evaluations to compare proprietary and open-source models for HPC kernel generation, demonstrating that while both models can generate valid parallel code, their performance and correctness vary significantly across programming models and hardware targets~\cite{parallel_kernels_2}.
Both studies highlighted persistent challenges, including missing synchronization, incorrect data scoping, and performance portability issues.
HPC-Coder~\cite{HPC-Coder} investigated \ac{llm}-based modeling of parallel programs, reporting mixed success in generating correct OpenMP pragmas and MPI calls, underscoring the difficulty of synthesizing semantically correct parallel code.
Godoy et al.~\cite{parallel_kernel_3} extended earlier evaluations by assessing \ac{llm}-generated parallel kernels across C\texttt{++}, Fortran, Python, and Julia, demonstrating that \acp{llm} can assist in auto-parallelization but still require expert validation to ensure correctness and performance.

While existing work focuses on code generation for different programming languages, GPU kernels, and fork-join parallelization, asynchronous task-based code generation remains largely unexplored.
Our work fills this existing research gap by evaluating task-based code generation using multiple frameworks across several \acp{llm} not only regarding compilation and correctness but also parallel scaling. 
Although crucial, parallel strong and weak scaling was not investigated in previous work.

\section{Methodology}\label{sec:methods}

We evaluate the ability of \acp{llm} to generate task-based parallel code using a set of benchmark problems that vary in algorithmic structure, dependency complexity, and sensitivity to compute and memory bandwidth limits.
The selected benchmarks span embarrassingly parallel workloads, divide-and-conquer algorithms, regular compute kernels, and iterative methods with global synchronization barriers:
\begin{enumerate}
    \item \textbf{Stochastic Approximation of $\pi$}, representing independent task execution with minimal synchronization.
    \item \textbf{Merge Sort}, exhibiting recursive task creation and explicit dependency constraints.
    \item \textbf{Matrix-Matrix Multiplication}, a dense linear algebra kernel with high arithmetic intensity and structured data access.
    \item \textbf{Conjugate Gradient Method}, an iterative solver involving sparse computations, reductions, and strict iteration ordering.
\end{enumerate}

Together, these benchmarks stress different aspects of task-based parallelization, including task granularity, dependency inference, and synchronization placement.
For each benchmark, we generate parallel implementations from three prompt modalities of progressively increasing specificity.
These modalities assess how input structure influences the model’s ability to infer parallelism:
\begin{enumerate}
    \item \textbf{Problem Description}, providing only a function signature, a high-level natural language specification, and the target parallel framework.
    \item \textbf{Sequential Code}, supplying a complete sequential C\texttt{++} implementation to be parallelized.
    \item \textbf{Parallel Pseudo Code}, augmenting the description with a structured parallel algorithm outline.

\end{enumerate}

This progression reflects common \ac{llm} usage scenarios, from algorithm design assistance to automated refactoring of existing code.
Each problem–prompt combination is evaluated across three \acp{llm} to enable comparative analysis:
\begin{enumerate}
    \item \textbf{ChatGPT-5}
    \item \textbf{Qwen-Coder 3 (30B)} \footnote{https://huggingface.co/Qwen/Qwen3-Coder-30B-A3B-Instruct}
    \item \textbf{Gemini-3}
\end{enumerate}
Our model selection covers different parameter scales, training objectives, and availability models, including both open-source and proprietary systems.

We evaluate \ac{llm}-generated task-based parallel code using three parallel programming frameworks that represent different abstraction levels and execution models:
\begin{enumerate}
    \item \textbf{OpenMP Tasking} is a widely used directive-based framework for shared-memory parallelism. Its explicit task creation and synchronization directives make it a natural baseline for evaluating an \ac{llm}’s ability to reason about dependencies and task granularity.
    \item \textbf{C\texttt{++} Standard Parallelism} provides high-level, library-based parallel abstractions such as parallel algorithms and execution policies. It allows us to assess whether \acp{llm} can exploit declarative parallelism. 
    \item \textbf{HPX} is a fully asynchronous, task-based runtime leveraging futures and dataflow semantics. We include HPX to evaluate \ac{llm} performance on fine-grained, dependency-driven parallelism. 
\end{enumerate}

Together, these frameworks cover a spectrum from directive-based to runtime-driven parallelism, enabling a comparative analysis of how \acp{llm} adapt to different task-based programming models.
For every combination of benchmark problem, prompt modality, \ac{llm}, and parallelization framework, we generate multiple code snippets.
Generated implementations are evaluated along three axes.
Correctness is assessed using problem-specific validation and reference outputs.
Complexity is quantified using lines of code, comment density, and control-flow patterns in the generated programs.
Scalability is evaluated by varying available parallel resources and analyzing speedup behavior.

This methodology isolates the impact of problem complexity, prompt modality, and model choice on the quality of \ac{llm}-generated task-based parallel code. 

\subsection{Evaluation Metrics}\label{sec:metrics}
We present \emph{pass@k} to analyze the correctness of the produced code, scc for complexity evaluation, and \emph{PCGQS} for benchmarking generated parallel code.

\subsubsection{Correctness}
The~\emph{pass@k} metric is widely used for evaluating \acp{llm} and their generated code~\cite{passatk}.
It quantifies the probability that at least one out of $k$ generated samples is correct.
The computation of~\emph{pass@k} is given by \Cref{eq:passatk}:

\begin{equation}
\text{pass@k} = 1 - \frac{\binom{n - c}{k}}{\binom{n}{k}}\label{eq:passatk}
\end{equation}

where $n$ denotes the total number of generated samples and $c$ the number of correct ones.
To assess the quality of the generated code, we introduce a set of levels that characterize the difficulty of correcting code errors.
We do not explicitly characterize by the type of error, i.e., compilation, runtime, or general algorithmic error.
However, typically compilation errors are easier to correct than runtime or algorithmic errors.
By categorizing the effort required to correct the code for successful use, this framework enables a more comprehensive evaluation of programs generated by \acp{llm}.
The set comprises the following levels:

\begin{equation}
C(c) =
\begin{cases}
1, & \text{No Fix – Fully correct, no modifications required}, \\
0.75, & \text{Easy – Minor adjustments are sufficient to achieve correctness}, \\
0.5, & \text{Medium – Moderate debugging, code modifications are required}, \\
0.25, & \text{Hard – Significant intervention is necessary to fix the code}, \\
0, & \text{Infeasible – The code cannot be feasibly corrected}.
\end{cases}
\label{eq:fix_levels}
\end{equation}

\subsubsection{Complexity}

Beyond executability and performance of the code, we also assess the intrinsic properties of the generated code examples.
We employ the tool \textit{scc}~\footnote{https://github.com/boyter/scc} to quantify the number of effective lines of code as well as the proportion of comment lines.
In addition, \textit{scc} provides a complexity metric approximating the cyclomatic complexity~\cite{cylomatic_complexity}, which can be used as an indicator of the expected maintenance effort associated with the code.

\subsubsection{Scaling}
To evaluate the ability of \acp{llm} to generate efficient parallel code, we define the~\emph{Parallel Code Generation Quality Score (PCGQS)}.
The metric combines functional correctness with empirical scaling behavior.
For a generated implementation $c$, the PCGQS is defined as:
\begin{equation}
\label{eq:PCGQS}
\mathrm{PCGQS}(c) = \frac{1}{2} C(c) + \frac{1}{2} S(c),
\end{equation}
where $C(c)$ denotes functional correctness as defined in \Cref{eq:fix_levels} and $S(c)$ captures parallel scalability.

Scaling behavior is evaluated using both strong and weak scaling experiments as defined in \Cref{eq:avg_strong} and \Cref{eq:avg_weak}, respectively.
$T_1$ is the single-core runtime, $T_p$ is the runtime using $p$ processing elements, and $\mathcal{P}$ is the set of evaluated core counts.
We combine strong and weak scaling into a single score as in \Cref{eq:mean_strong_weak}
\begin{align}
S_{\text{strong}}(c) &=
\frac{1}{|\mathcal{P}|}
\sum_{p \in \mathcal{P}}
\frac{T_1}{p \cdot T_p}\label{eq:avg_strong}\\
S_{\text{weak}}(c) &=
\frac{1}{|\mathcal{P}|}
\sum_{p \in \mathcal{P}}
\frac{T_1}{T_p}\label{eq:avg_weak}\\
S(c) &= \frac{1}{2} S_{\text{strong}}(c) + \frac{1}{2} S_{\text{weak}}(c)
\label{eq:mean_strong_weak}
\end{align}

\section{Results}\label{sec:results}

To ensure experimental reproducibility, all experiments are conducted within a fixed testing environment. The experimental setup consisted of a system equipped with a dual-socket AMD EPYC 7742 CPU. For the scaling experiments, we evaluate the code on up to 128 threads.
We bind the thread to CPU cores to eliminate potential unseen effects introduced by multi-threading. Based on a preliminary task-scaling analysis, the number of tasks is fixed at 128 for all experiments. Regarding the key software specifications, we use GCC version $13.3.0$, OpenMP version 4.5, and HPX version $1.11.0$.
We begin with an assessment of code correctness, followed by an analysis of complexity.
Finally, we examine the scaling behavior.

\subsection{Correctness}

\begin{figure}[ht]
    \centering
    \begin{subfigure}[t]{0.48\textwidth}
        \centering
        \includegraphics[width=0.99\linewidth]{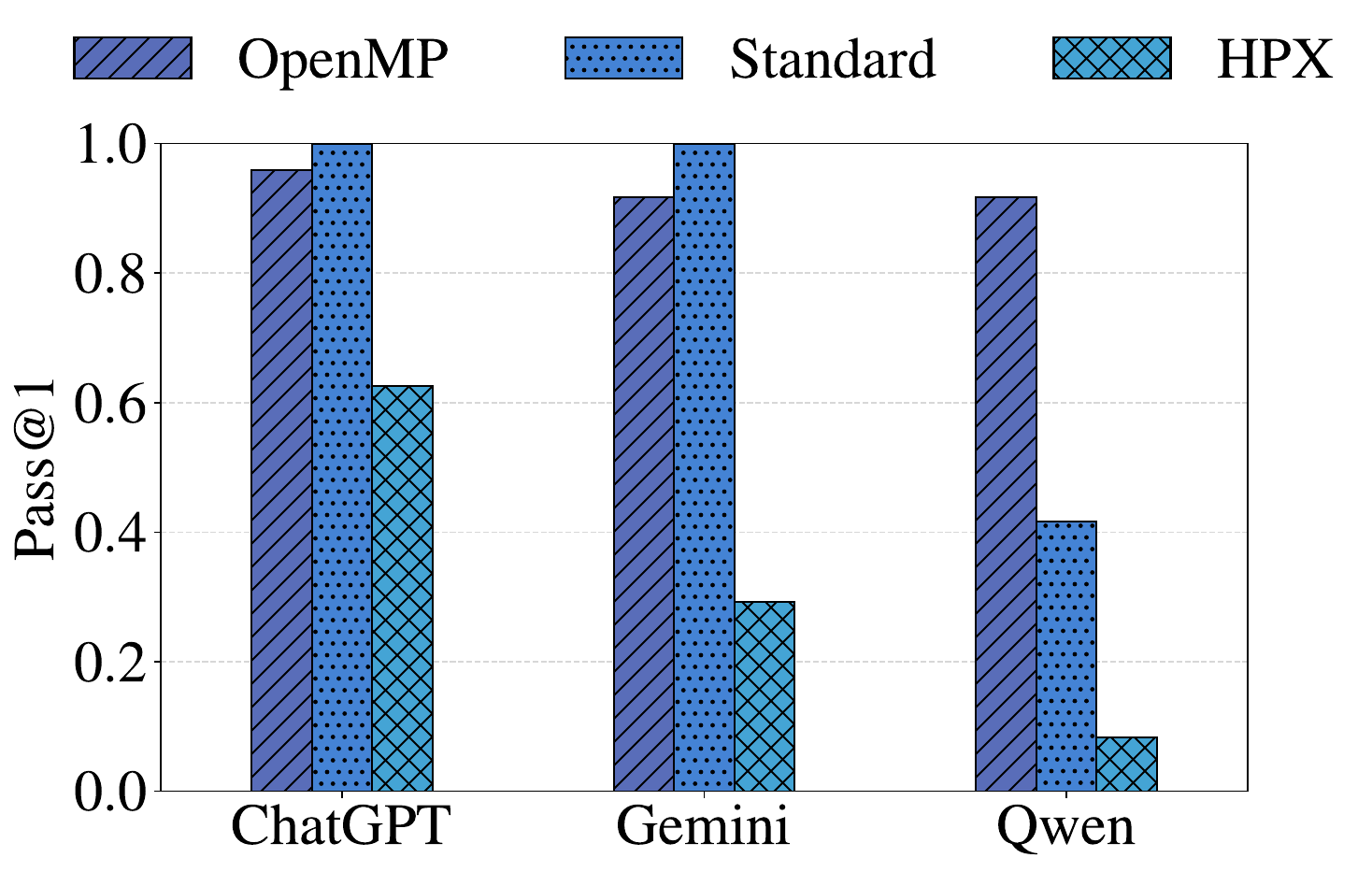}
        \caption{Grouped by framework.}
        \label{fig:pass_at_1_no_correction_framework}
    \end{subfigure}
    \begin{subfigure}[t]{0.48\textwidth}
        \centering
        \includegraphics[width=0.99\linewidth]{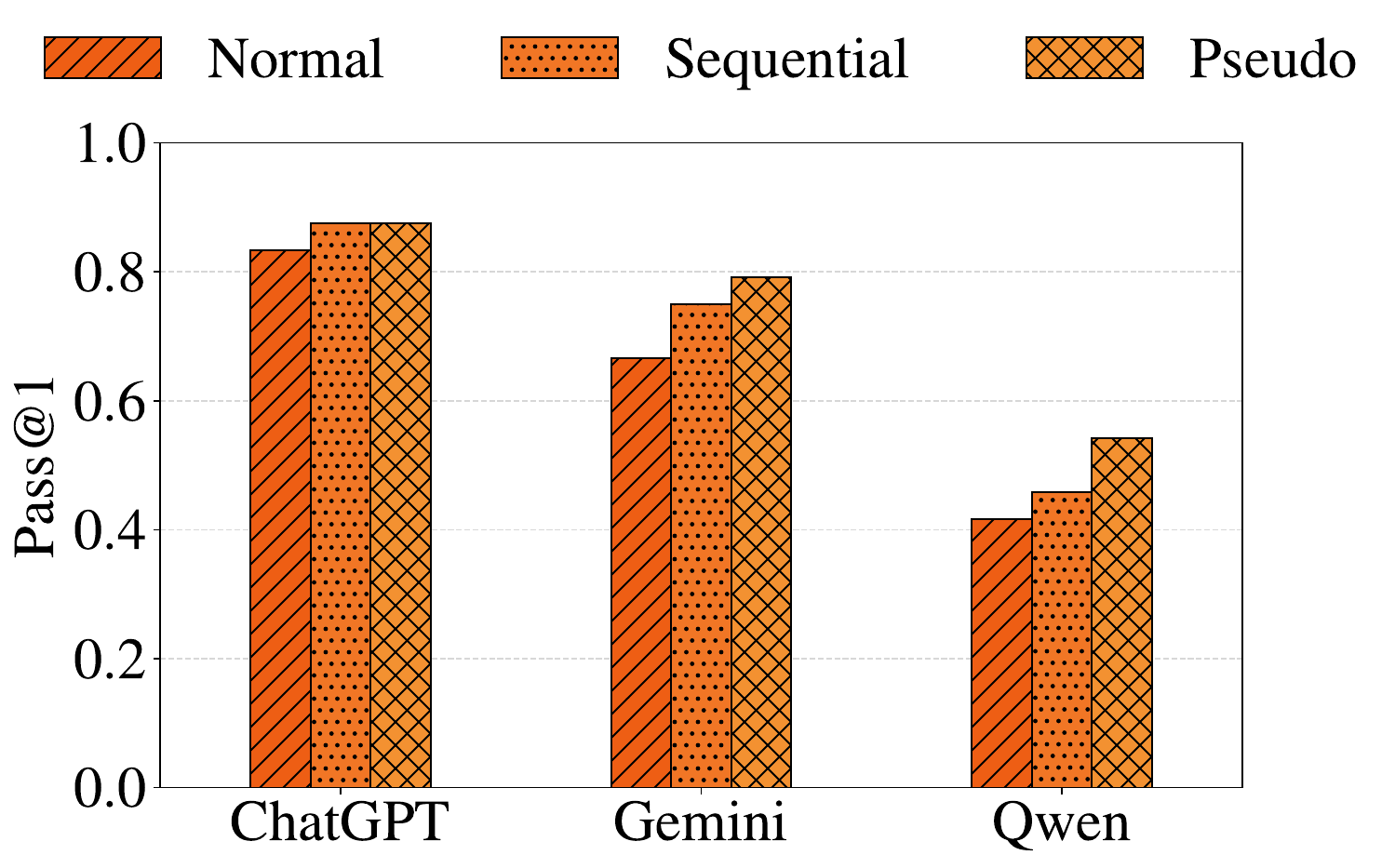}
        \caption{Grouped by prompt type.}
        \label{fig:pass_at_1_no_correction_prompt}
    \end{subfigure}
    \begin{subfigure}[t]{0.48\textwidth}
        \centering
        \includegraphics[width=0.99\linewidth]{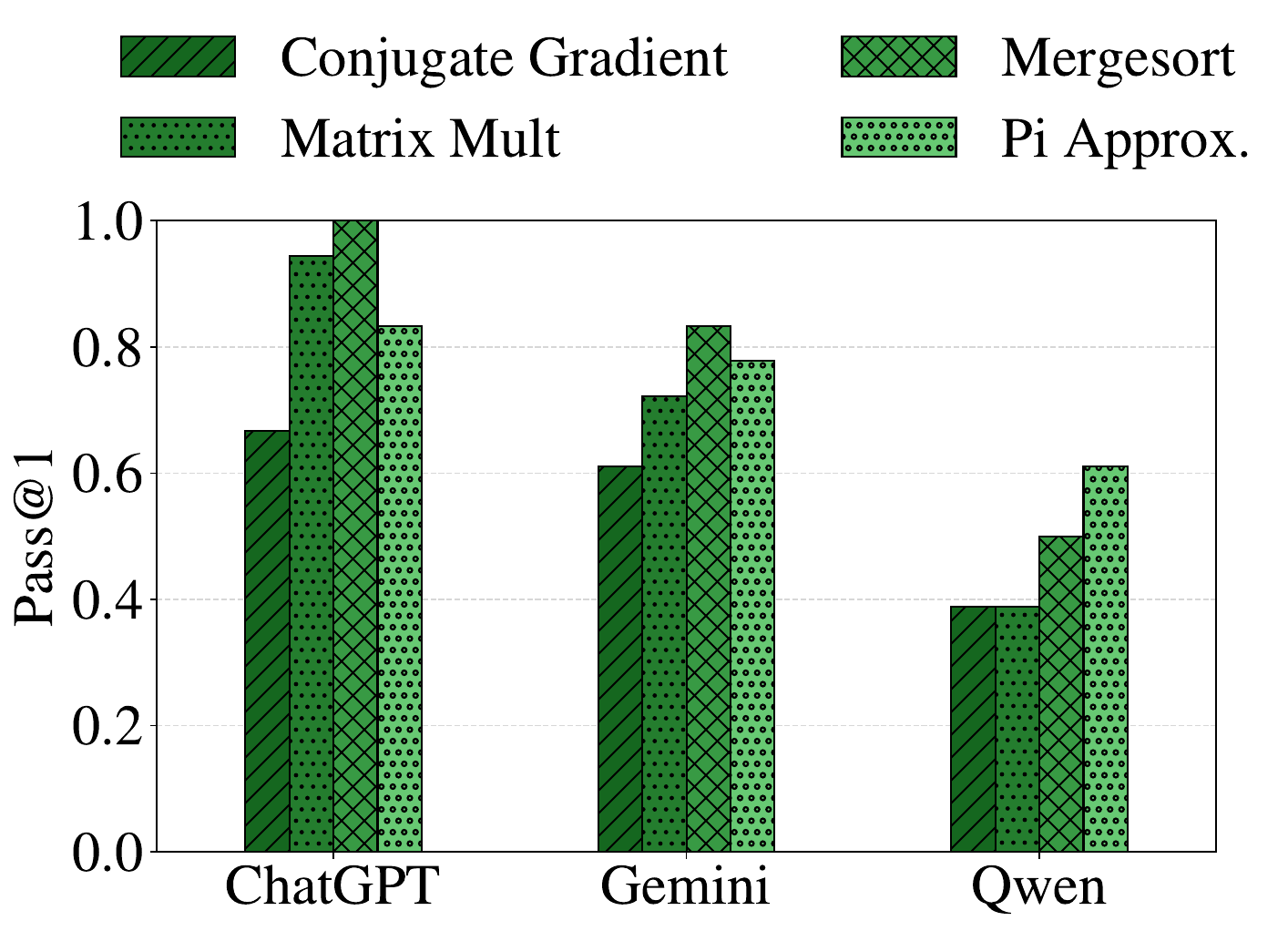}
        \caption{Grouped by problem.}
        \label{fig:pass_at_1_no_correction_problem}
    \end{subfigure}
    \caption{Pass@1 metric for the respective \acp{llm} without any correction.}
    \label{fig:pass_at_1_no_correction}
\end{figure}

We begin by analyzing the out-of-the-box performance of the three \acp{llm}. For each \ac{llm}, we compute the pass@1 metric across all remaining experimental groups.
In this way, \Cref{fig:pass_at_1_no_correction} provides an initial, yet meaningful, overview of model performance.
Even in these first results, substantial differences between the models are apparent.
ChatGPT exhibits the strongest out-of-the-box performance in terms of generating correct code.
Google's Gemini ranks second, while the open-source model Qwen-Coder trails at a considerable distance.
As shown in \Cref{fig:pass_at_1_no_correction_framework}, OpenMP performs robustly across all models.
C\texttt{++} standard parallelism achieves near-perfect results for ChatGPT and Gemini but performs very poorly for Qwen-Coder.
HPX, by contrast, starts at a low pass@1 rate and degrades rapidly across all models.
\Cref{fig:pass_at_1_no_correction_prompt} reveals a similar overall pattern with respect to the different prompt formulations.
No statistically significant performance differences can be attributed to the prompt type. Prompts containing parallel pseudocode yield only marginal improvements.
When examining the individual benchmark problems in \Cref{fig:pass_at_1_no_correction_problem}, several noteworthy observations can be made.
Although the parallel approximation of $\pi$ might intuitively appear simpler than parallel mergesort, both ChatGPT and Gemini generate more reliable code for mergesort.
This is likely due to mergesort being a canonical programming exercise commonly encountered in introductory computer science courses, resulting in a large number of readily available implementations.
Finally, the \ac{cg} benchmark proves sufficiently challenging that the pass@1 metric drops markedly in comparison to the generally better-performing matrix–matrix multiplication benchmark.

\begin{figure}[ht]
    \centering
    \begin{subfigure}[t]{0.48\textwidth}
        \centering
        \includegraphics[width=0.99\linewidth]{figures/model_benchmark_no_correction.pdf}
        \caption{No fix}
        \label{fig:pass_at_1_no}
    \end{subfigure}
    \begin{subfigure}[t]{0.48\textwidth}
        \centering
        \includegraphics[width=0.99\linewidth]{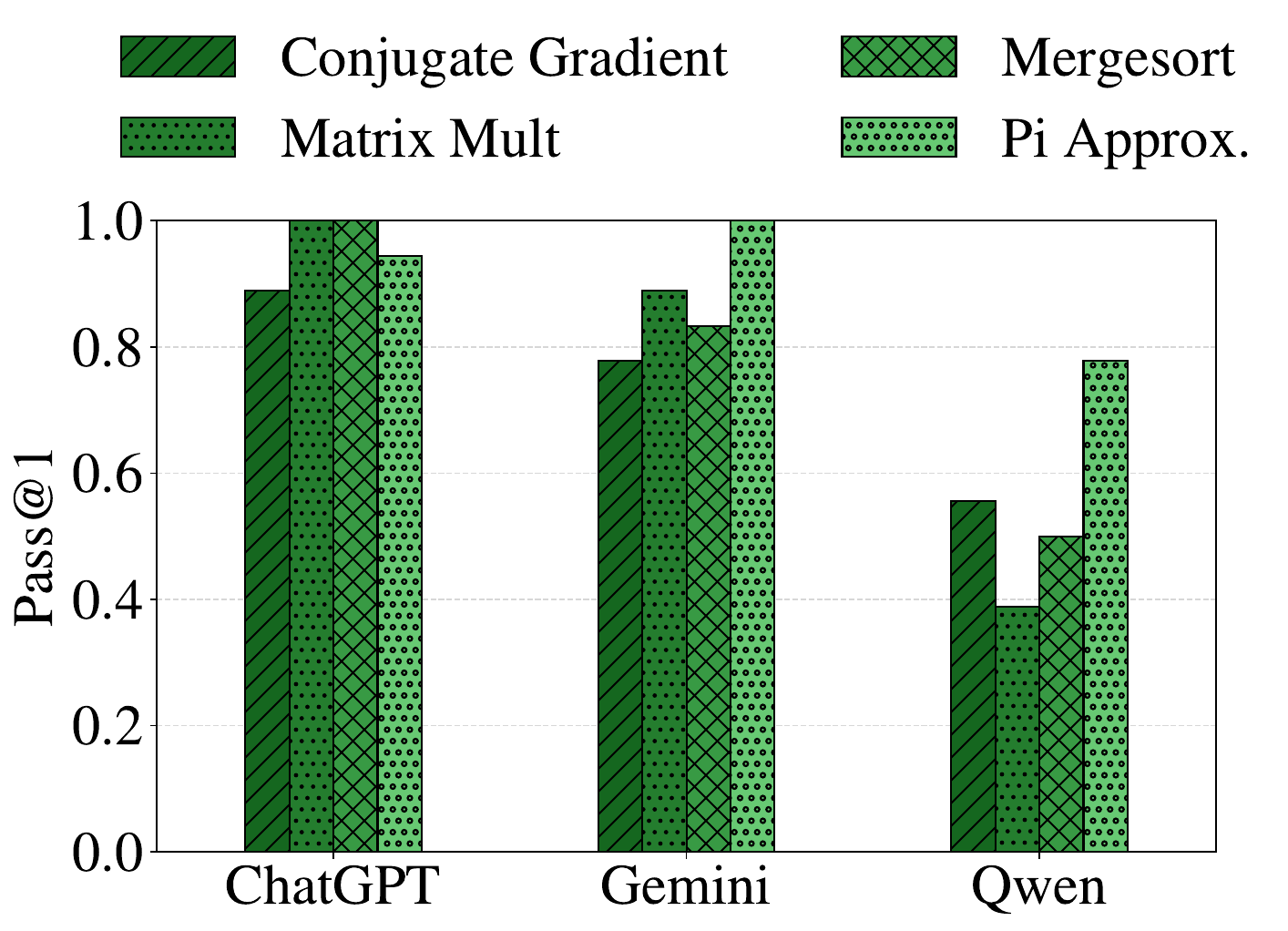}
        \caption{Easy}
        \label{fig:pass_at_1_easy}
    \end{subfigure}
    \begin{subfigure}[t]{0.48\textwidth}
        \centering
        \includegraphics[width=0.99\linewidth]{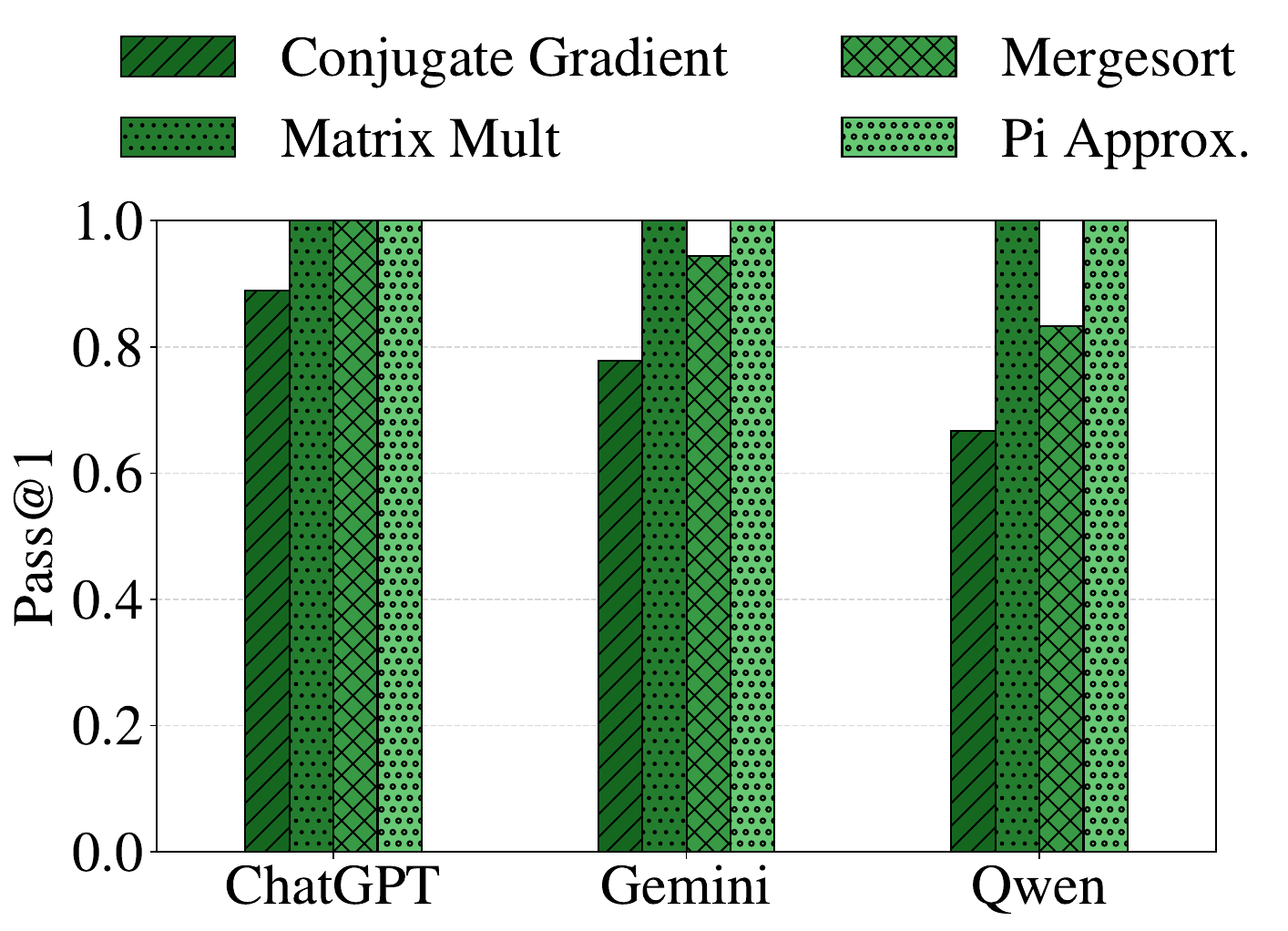}
        \caption{Medium}
        \label{fig:pass_at_1_medium}
    \end{subfigure}
    \begin{subfigure}[t]{0.48\textwidth}
        \centering
        \includegraphics[width=0.99\linewidth]{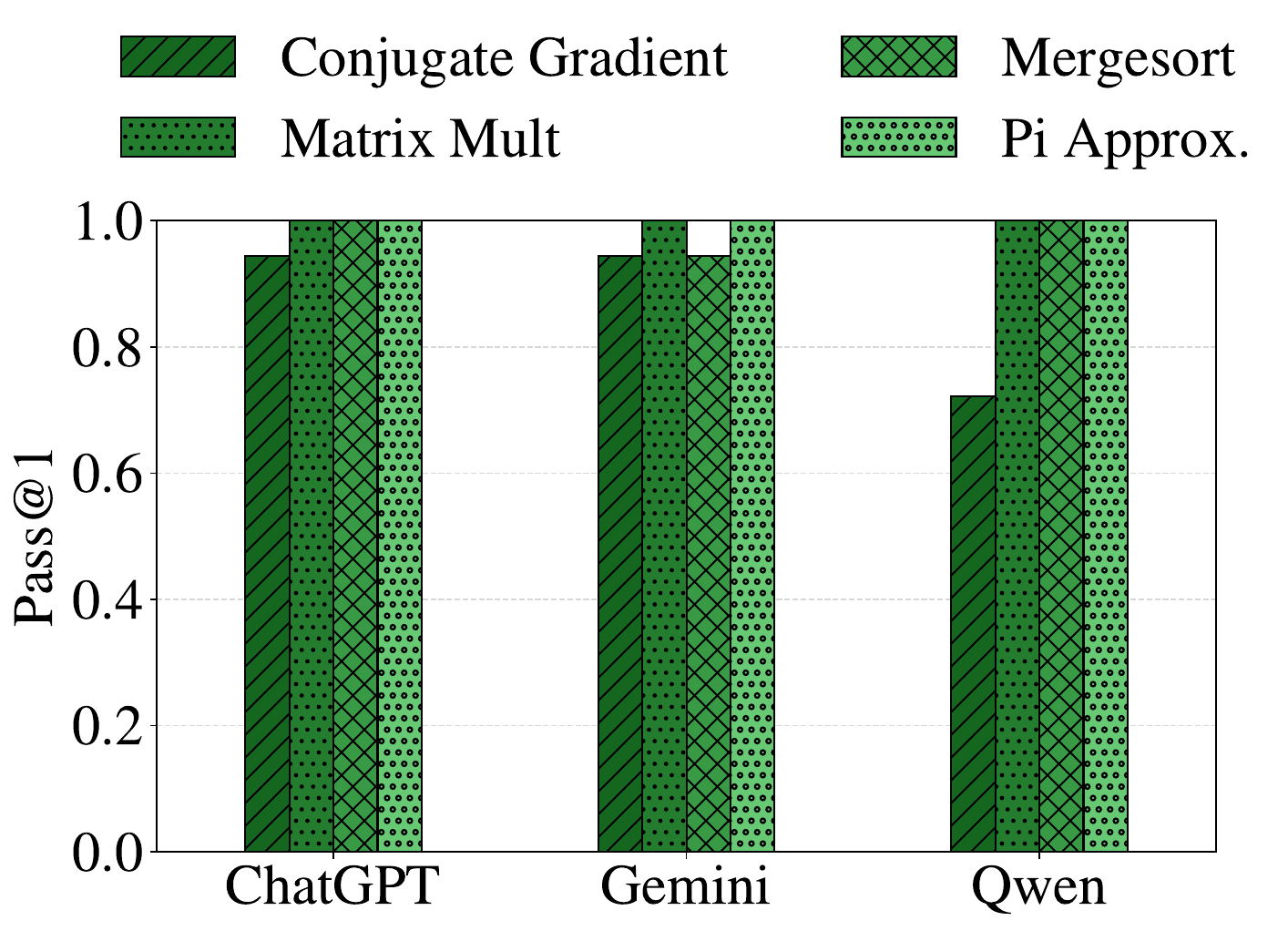}
        \caption{Hard}
        \label{fig:pass_at_1_hard}
    \end{subfigure}
    \caption{Pass@1 metric for the level of fixes, such that the program is correct and runs}
    \label{fig:pass_at_1_fixes}
\end{figure}

In \Cref{fig:pass_at_1_fixes}, we analyze the performance improvements obtained when limited human corrections are permitted.
The required modifications to obtain correct code are categorized into the five fix levels introduced in \Cref{sec:metrics}.
As a baseline, \Cref{fig:pass_at_1_no} reports the pass@1 metric without any human intervention.
This plot is identical to \Cref{fig:pass_at_1_no_correction_problem}.
Allowing only easy fixes already leads to a noticeable increase in the proportion of functioning code, as shown in \Cref{fig:pass_at_1_easy}.
Minor corrections are sufficient to resolve a substantial fraction of failures across all models and benchmark problems.
When medium-level fixes are permitted, the pass@1 metric reaches 100\% for all models on several benchmarks.
Notably, Qwen-Coder surpasses Gemini in this setting by achieving perfect results on both matrix-matrix multiplication and the approximation of~$\pi$.
In contrast, mergesort, which exhibited the highest pass@1 rate without correction, falls behind matrix-matrix multiplication and $pi$ approximation at the medium fix level.
With hard fixes, a pass@1 rate close to 100\% is achieved for nearly all model-benchmark combinations.
The only notable exception is the \ac{cg} benchmark, which continues to show significantly lower pass@1 values for Qwen-Coder.
In the following, we describe the concrete code modifications applied and justify their assignment to the respective fix categories.

When it comes to the type of errors, the most common header-related errors include missing headers, outdated headers, or the use of non-existent headers.
We categorize them as easy fixes.
For HPX and C\texttt{++} standard parallelism, lambda expressions frequently exhibit incorrect capture clauses.
These are medium fixes.
A recurring error class, specific to generated  HPX-Code, consists of outdated or incorrect namespaces, such as the use of \texttt{hpx::parallel::for\_loop} instead of \texttt{hpx::experimental::for\_loop}.
Among genuine coding errors, incorrect handling of futures in HPX is particularly prevalent, most often caused by misplaced or missing \texttt{.get()} calls.
More severe errors arise from incorrect sharing of arrays across parallel regions, which frequently leads to segmentation faults or silent data corruption.
We additionally observe failures when executing the generated code on odd core counts, specifically for 24, 48, and 96 cores.
If the code fails in any of the scaling experiments and the required fix is infeasible, we classify the corresponding output as non-executable.

\subsection{Complexity}
\Cref{tab:complexity} presents a comparison of the influence of different \acp{llm}, programming frameworks, and prompting strategies on the structure and complexity of the generated code. 
The number of lines of code stays constant for all categories.
Notable differences are observed in the extent of code documentation: Gemini-generated code contains a higher density of comments, whereas ChatGPT produces the fewest comments. 
Furthermore, the level of abstraction in the prompt is directly correlated with the complexity of the resulting code. 
Prompts that integrate pseudo code provide a clear structure, facilitating code generation. 
In contrast, prompts that provide a sequential implementation encourage the model to adhere closely to that structure, with parallel constructs added around it. 
This approach results in increased structural complexity in the generated code.

\begin{table}[!t]
\centering
\newcolumntype{Y}{>{\centering\arraybackslash}X} 

\begin{tabularx}{\textwidth}{lYYYY}
\toprule
\textbf{Category} & \textbf{\#Lines} & \textbf{\#Code} & \textbf{\#Comments} & \textbf{Complexity} \\
\midrule
\textit{LLM} & & & & \\
ChatGPT & 67.25 & 51.27 & 4.66 & 8.65 \\
Gemini & 75.97 & 51.07 & 12.80 & 9.06 \\
Qwen & 71.89 & 50.78 & 8.63 & 9.66 \\
\midrule
\textit{Framework} & & & & \\
OpenMP & 70.11 & 51.14 & 8.16 & 9.06 \\
Standard C\texttt{++}& 70.93 & 50.26 & 8.71 & 10.26 \\
HPX & 74.12 & 51.75 & 9.25 & 7.99 \\
\midrule
\textit{Prompt Level} & & & & \\
Normal & 73.46 & 50.43 & 10.35 & 9.01 \\
Sequential & 73.86 & 52.70 & 8.86 & 10.23 \\
Pseudo & 67.69 & 49.97 & 6.87 & 8.04 \\
\bottomrule
\end{tabularx}
\caption{Comparison of generated code with regards to length and complexity.}
\label{tab:complexity}
\end{table}

\subsection{Scaling}
We next analyze the scaling behavior of the generated code using the strong and weak scaling definitions given in \Cref{eq:avg_weak,eq:avg_strong}.
For that, we fixed all errors, including the ones categorized as hard.
Non-executable code is included as zero scaling.

\begin{figure}[ht]
    \centering
    \begin{subfigure}[t]{0.48\textwidth}
        \centering
        \includegraphics[width=0.99\linewidth]{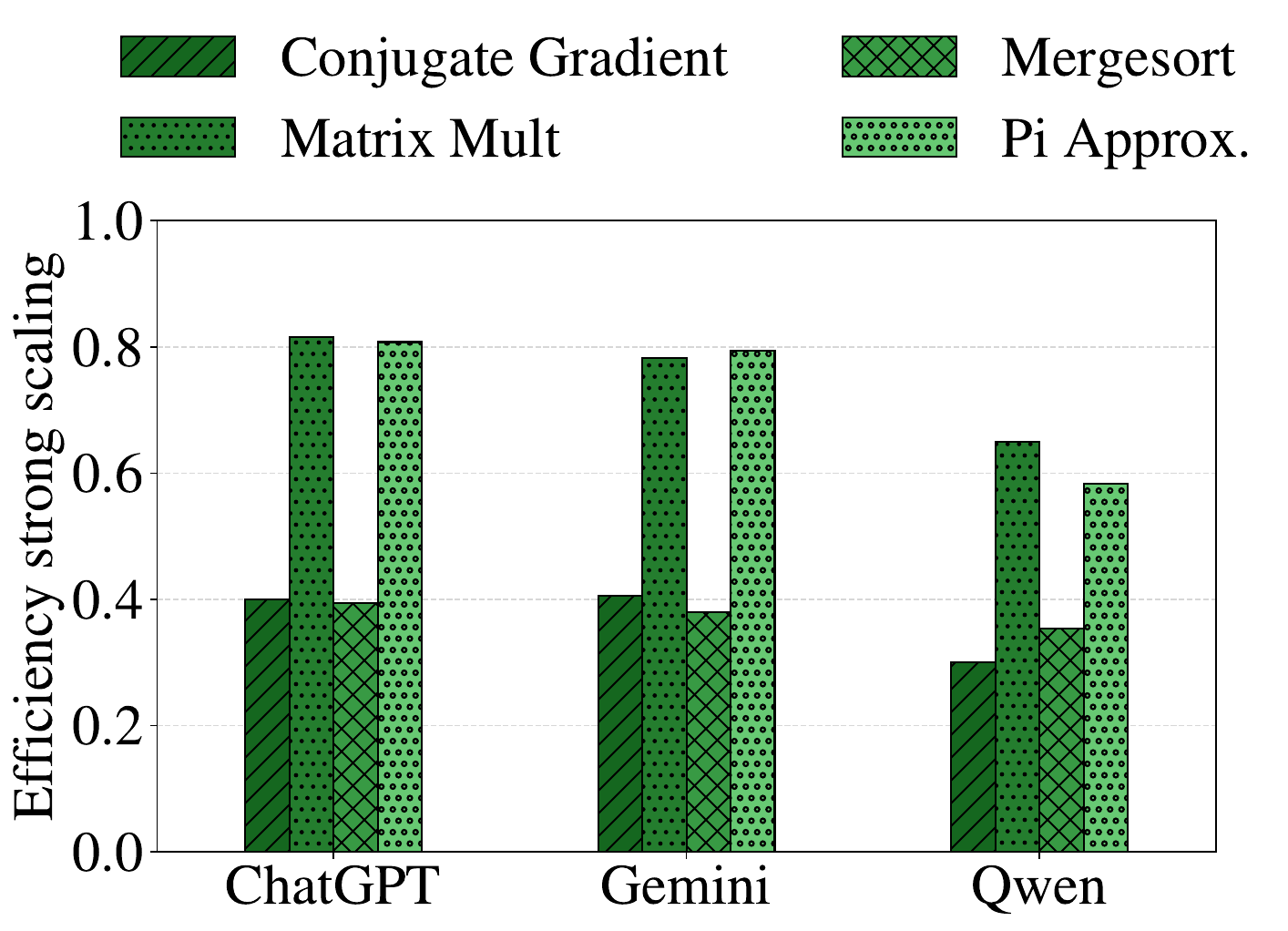}
        \caption{Strong scaling}
        \label{fig:strong_scaling}
    \end{subfigure}
    \begin{subfigure}[t]{0.48\textwidth}
        \centering
        \includegraphics[width=0.99\linewidth]{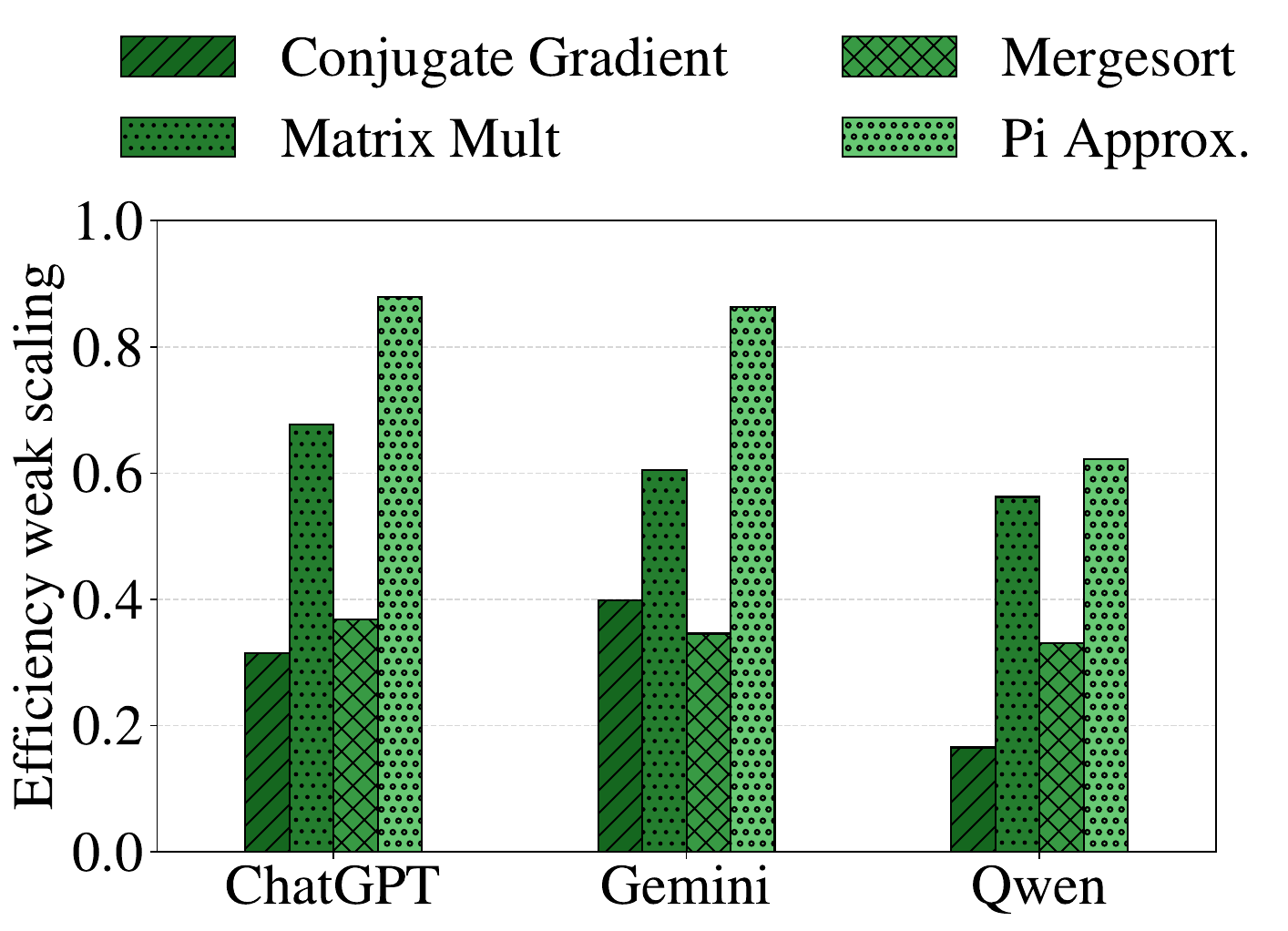}
        \caption{Weak scaling}
        \label{fig:weak_scaling}
    \end{subfigure}
    \caption{Scaling of the generated code grouped by benchmark problem.}
    \label{fig:scaling}
\end{figure}

The resulting scaling performance is summarized in \Cref{fig:scaling}.
Several noteworthy observations emerge from this analysis.
As expected, the approximation of~$\pi$ and matrix-matrix multiplication demonstrate consistently superior scaling behavior compared to mergesort and the \ac{cg} algorithm.
This difference arises because both the $\pi$ approximation and matrix-matrix multiplication exhibit regular memory access patterns.
They also require minimal synchronization between parallel tasks.
In contrast, mergesort and \ac{cg} involve irregular control flow, recursive or iterative dependencies, and frequent reductions or synchronization points.
These characteristics limit the achievable speedup as the number of cores increases.
While the code generated by ChatGPT and Gemini exhibits very similar scaling trends and generally maintains performance across increasing core counts, Qwen-Coder’s implementations scale considerably worse.
This poor scaling is particularly evident in the approximation of~$\pi$, which in theory should achieve near-linear speedup with increasing cores due to its inherently parallel structure.
A major factor behind this limitation is that Qwen-Coder frequently implements the HPX variants using a single global random number generator.
Consequently, all concurrent tasks, including up to 128 simultaneous threads in our experiments, must access the same generator for random values.

\begin{figure}[!t]
    \centering
    \begin{subfigure}[t]{0.48\textwidth}
        \centering
        \includegraphics[width=0.99\linewidth]{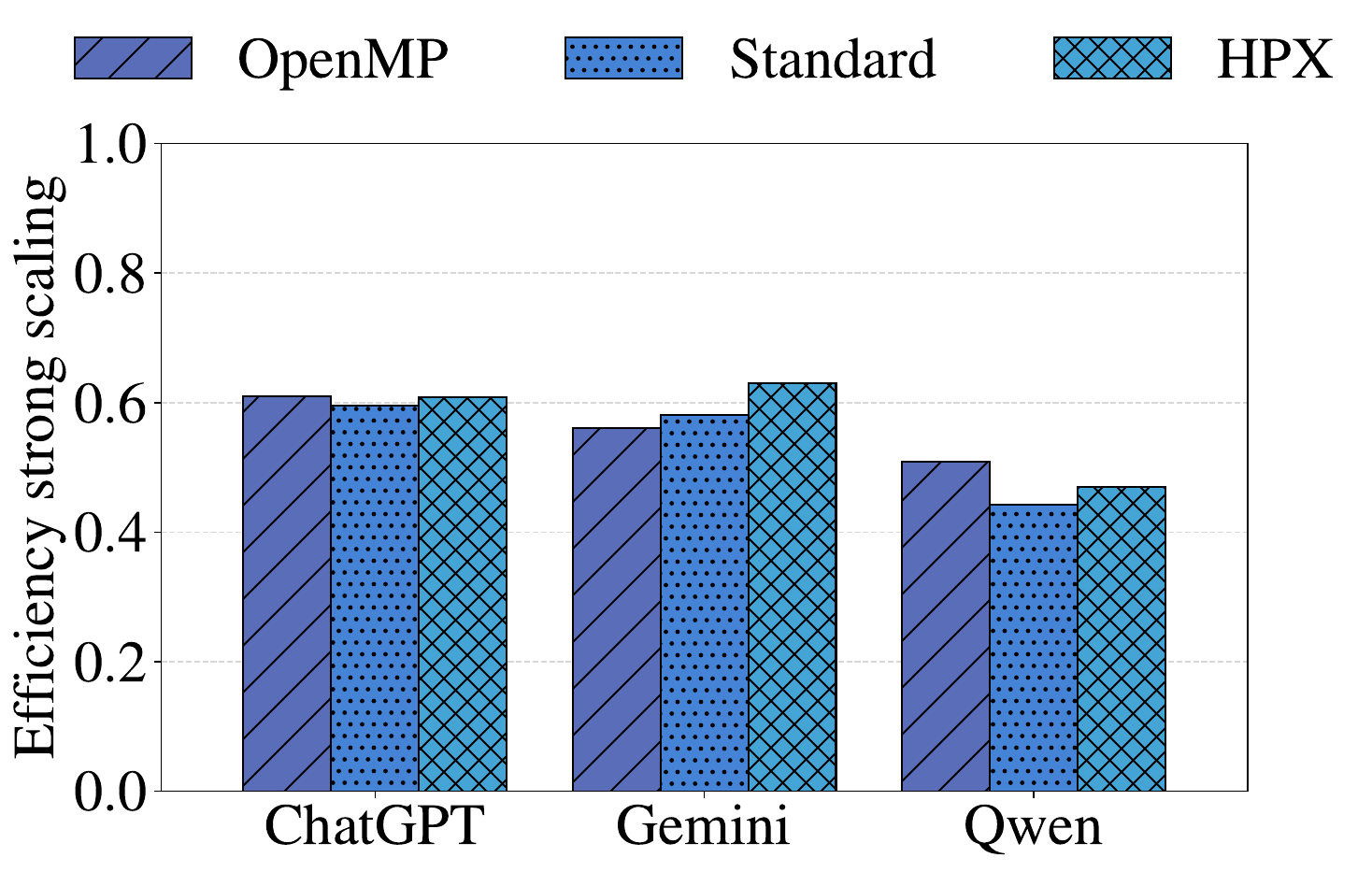}
        \caption{Strong scaling}
        \label{fig:framework_strong_scaling}
    \end{subfigure}
        \begin{subfigure}[t]{0.48\textwidth}
        \centering
        \includegraphics[width=0.99\linewidth]{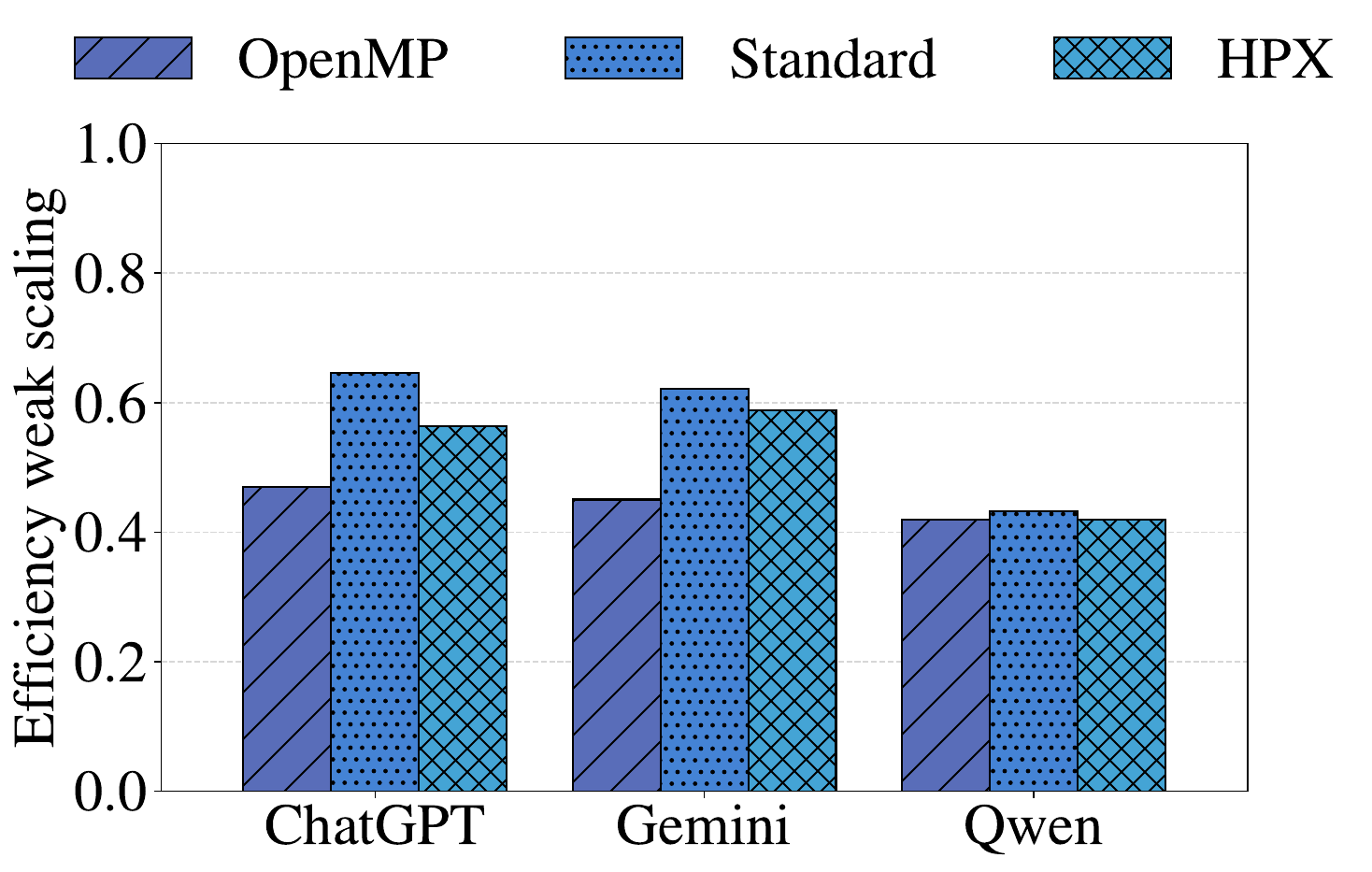}
        \caption{Weak scaling}
        \label{fig:framework_weak_scaling}
    \end{subfigure}
    \caption{Scaling of the generated code grouped by framework.}
    \label{fig:framework_scaling}
\end{figure}

When we categorize the results by the parallelization framework rather than by the benchmark problem, we can identify the best combinations of framework and model.
The data in \Cref{fig:framework_scaling} indicates that C\texttt{++} standard parallelism consistently achieves the highest weak scaling efficiency across all models.
In contrast, OpenMP delivers the worst efficiency.
Generated OpenMP code often confuses task-based parallelism with classic loop-based parallelism, for example, by mixing \texttt{single} or \texttt{parallel} regions with \texttt{parallel for} constructs incorrectly.

While we see larger discrepancies between the frameworks for weak scaling, the strong scaling performance is very similar across all frameworks.
Only code generated by Qwen-Coder achieves slightly lower efficiency.
The results suggest that C\texttt{++} standard parallelism is the most robust choice for code generated by the evaluated \acp{llm}, while HPX can be competitive if the model correctly handles framework-specific details.
Finally, 

Last but not least, we use the PCGQS-Score from \Cref{eq:PCGQS} to evaluate the tested \acp{llm}. \Cref{tab:model_scores} shows the combined score over all experiments, frameworks, and prompt levels.
ChatGPT demonstrates the strongest overall performance, whereas the open-source model Qwen exhibits substantial difficulties in terms of response correctness, which markedly reduced its combined score. 

\begin{table}[h]
  \centering
  \begin{tabular}{cccc}
  \toprule
  &ChatGPT&Gemini&Qwen\\
  \midrule
  PCGQS-Score&0.7425&0.7008&0.5702 \\
  \bottomrule
  \end{tabular}
  \caption{PCGQS-Score for the evaluated \acp{llm}.}
  \label{tab:model_scores}
\end{table}

\section{Conclusion and Outlook}\label{sec:conclusion}

In this work, we systematically evaluated the ability of \acp{llm} to generate task-based parallel code across multiple benchmarks with varying algorithmic structure and dependency complexity.  
Our results show that \acp{llm} can reliably handle embarrassingly parallel problems and simple task graphs, indicating a solid grasp of basic parallel abstraction.  
For more complex patterns, such as iterative algorithms with intrinsic synchronization, the models frequently produce suboptimal or incorrect task dependencies.  
These shortcomings manifest as missing synchronization or latent race conditions, despite syntactically correct code.  
The proposed evaluation metrics, including the PCGQS, help distinguish models beyond functional correctness by capturing structural parallelism and scalability potential.  

Regarding future work, we plan to extend the benchmark suite toward larger, more irregular, and multi-stage parallel applications.  
Further improvements may be achieved through refined prompting, explicit dependency representations, and tool-assisted generation workflows.  
Together, these directions aim to improve both the reliability and practical usefulness of \acp{llm} for asynchronous task-based parallel programming.
Lastly, recent advances in agentic AI have led to substantial improvements in code generation performance, needing systematic investigation in the context of asynchronous task-based parallelization frameworks.

\subsection*{AI Use Disclosure}

Generative AI tools, including Grammarly~\cite{grammarly}, DeepL~\cite{deepl}, Gemini~\cite{gemini}, and ChatGPT~\cite{chatgpt}, were employed to enhance the clarity, grammar, and overall coherence of the manuscript. All technical content, data analyses, and research findings were conceived and developed independently by the authors. AI-assisted outputs were carefully reviewed, verified, and edited by the authors to ensure factual accuracy, interpretive rigor, and scholarly integrity. The final manuscript reflects the authors’ original intellectual contributions and analytical work.
\bibliographystyle{ieeetr}
\bibliography{main}

\end{document}